\documentclass[preprint2]{aastex}
\newcommand{\myemail}{maggie@physics.mcgill.ca}
\newcommand{\psr}{PSR~J0205+6449}

\newcommand{\rxte}{{\textit{RXTE}}}

\newcommand{\nudotdotdot}{{\ifmmode\stackrel{\bf \,...}{\textstyle \nu}\else$\stackrel{\,\...}{textstyle \nu}$\fi}}
\newcommand{\degrees}{^{\circ}}

\newcommand{\xte}{{\textit {RXTE}}}
\newcommand{\cxo}{\textit{Chandra X-ray Observatory}}

\usepackage{graphicx}

\shorttitle{Timing the Pulsar in 3C~58}
\shortauthors{Livingstone et al.}

\begin{document}
\title{X-ray and Radio Timing of the Pulsar in 3C~58}

\author{Margaret A.~Livingstone \altaffilmark{1}}
\affil{Department of Physics,
McGill University, Montr\'eal, Qu\'ebec, Canada, H3A 2T8}

\author{Scott M.~Ransom}
\affil{National Radio Astronomy Observatory (NRAO),
Charlottesville, VA 22903}

\author{Fernando Camilo}
\affil{Columbia Astrophysics Laboratory, Columbia University,
New York, NY 10027}

\author{Victoria M.~Kaspi}
\affil{Department of Physics, McGill University, Montr\'eal, Qu\'ebec, Canada, H3A 2T8}

\author{Andrew G.~Lyne}
\affil{University of Manchester, Jodrell Bank Centre for Astrophysics,
Manchester M13 9PL, UK}

\author{Michael Kramer}
\affil{University of Manchester, Jodrell Bank Centre for Astrophysics,
Manchester M13 9PL, UK}
\affil{Max-Planck Institut fuer Radioastronomie, 53121 Bonn, Germany}

\and

\author{Ingrid H.~Stairs}
\affil{Department of Physics of Astronomy, University of British Columbia,
Vancouver, British Columbia, Canada, V6T 1Z1}
\affil{Australia Telescope National Facility, CSIRO, Epping, NSW
1710, Australia}
\affil{Centre for Astrophysics \& Supercomputing, Swinburne University
of Technology, Hawthorn, VIC 3122, Australia}

\altaffiltext{1}{\myemail}

\clearpage

\begin{abstract}
We present timing data spanning 6.4\,yr for the young and
energetic \psr, in the
supernova remnant 3C~58. Data were obtained with the
\textit{Rossi X-ray Timing Explorer}, the Jodrell Bank Observatory and
the Green Bank Telescope. We present phase-coherent timing analyses 
showing timing noise and two spin-up glitches with fractional
frequency increases of 
$ \sim 3.4\times10^{-7}$ near MJD~52555, 
and $ \sim 3.8\times10^{-6}$ between MJDs~52777 and
53062. These glitches are unusually large if the pulsar was created in 
the historical supernova in 1181 as has been suggested.
For the X-ray timing we developed a new unbinned
maximum-likelihood method for determining pulse arrival times which
performs significantly better than the traditional binned techniques.
In addition, we present an X-ray pulse profile analysis of four years of
\xte\ data showing that the pulsar is detected up to $\sim$40\,keV.
We also present the first measurement of the phase offset between the
radio and X-ray pulse for this source, showing that the radio pulse leads
the X-ray pulse by $\phi=0.10\pm0.01$ in phase. We compile all 
known measurements of the phase offsets
between radio and X-ray and radio and $\gamma$-ray pulses for X-ray and
$\gamma$-ray pulsars.  We show that there
is no relationship between pulse period and phase offset, 
supported by our measurement of the phase offset for \psr. 

\end{abstract}

\keywords{pulsars: general---pulsars: individual (\objectname{\psr})---X-rays: stars}

\section{INTRODUCTION}
\label{sec:intro}

\psr\ is a 65-ms rotation-powered pulsar residing in the
center of the supernova remant 3C~58. The pulsar was discovered in
a 2002 \cxo\ ({\textit{CXO}}) observation of the source \citep{mss+02} and subsequently
detected as a radio pulsar \citep{csl+02}. It is one of the most 
energetic pulsars in the Galaxy, with a spin-down luminosity of
$\dot{E}\sim 2.7\times10^{37}$erg\,s$^{-1}$.

The possible association between the 3C~58 pulsar/pulsar wind nebula complex 
and the historical supernova SN 1181 has been a matter of debate. 
3C~58 is the only known source positionally coincident and
energetically compatible with the historical supernova,    
strongly suggesting an association \citep{sg02a}. However, in recent years,  
evidence has been mounting that the association          
may be spurious and that the true age of the pulsar may be closer to its 
characteristic age, $\tau_c \equiv \nu/2\dot{\nu}=5.4$\,kyr (where $\nu$ is
the pulsar's rotation frequency and $\dot{\nu}$ its derivative), rather than
the implied historic age of 828\,yr. If the source is 828\,yr-old, the 
size of the remnant implies a large expansion velocity
that is hard to reconcile with the measured velocities of the optical
filaments \citep{fes83} and the expansion speed of the synchrotron
bubble \citep{bkw01b, bie06}. The characteristic age estimate assumes that the pulsar
was born spinning rapidly ($\nu_0 \gg \nu$) and that the temporal spin evolution of
the pulsar has proceeded according to a simple magnetic dipole spin-down model.
These assumptions are known to fail in some cases, such as for 
PSR~J1811$-$1925 in the supernova remnant G11.2$-$0.3, for 
which the pulsar's characteristic age appears to be a factor of $\sim$15
greater than that of the remnant \citep[e.g.][]{ttd+99,krv+01,tr03}.
The two ages can be reconciled if PSR~J1811$-$1925 was born spinning
slowly, with $\nu_0 \simeq \nu$. Similarly, the pulsar J0538+2817 has a
characteristic age of 620\,kyr, but a well established kinematic age
of $\sim$40\,kyr, implying a long initial spin period of $\sim$138\,ms
\citep{klh+03,nrb+07}. Likewise, if \psr\ was born spinning
with a period of $\sim$60\,ms, the estimated age of the pulsar could
be reconciled with the historical
supernova age \citep{mss+02}. An alternate 
explanation of the age disparity for 
\psr\ is if the pulsar
were born with a short spin period but evolved more rapidly than is
typically assumed for magnetic dipole braking.

In principle, it may be possible to to gain
information about a pulsar's true age via its glitching behavior.
Pulsars in general, and young pulsars in particular, 
exhibit two types of rotational irregularities superposed
on their secular spin down. Glitches are characterized by 
a sudden increase in $\nu$,
sometimes followed by an exponential decay, and are often accompanied
by an increase in the magnitude of $\dot{\nu}$. Glitches are common
in young pulsars, particularly in pulsars aged $\sim$10-100\,kyr,
and provide valuable information about the superfluid interiors of pulsars. 
The fractional sizes of detected glitches span 5 orders of magnitude, with 
$\Delta{\nu}/{\nu} \sim 10^{-10} - 10^{-5}$ \citep{js06,hlj+02}.
The nature of glitches appears to change with age.
Glitches in the youngest Crab-like pulsars (1-2\,kyr) tend to be small in magnitude, 
$\Delta{\nu}/{\nu}\sim 10^{-9}-10^{-8}$, and
if the frequency recovers, it tends to recover almost completely. 
By contrast, glitches in older Vela-like pulsars are typically larger, with
magnitudes of $\Delta{\nu}/{\nu}\sim 10^{-7}-10^{-6}$, and typically recover only a
small percentage of the frequency change \citep{lsg00}.

Young pulsars are also prone to large stochastic variations in their 
spin-down rates, known as timing noise.
These noise processes are observed as long-term 
trends in
timing residuals after the removal of deterministic spin-down effects.
Timing noise tends to be the most severe in young pulsars and a
correlation with $\dot{\nu}$ is well established \citep{cd85,antt94,ulw06}.
The low-frequency and broad-band nature of
timing noise leads to difficulties in its analysis.
Moreover, the physical causes of timing noise are poorly understood.
It could be that fluctuations in the superfluid interiors of neutron
stars cause a torque on the pulsar \citep{cg81}, or that interactions
between the pulsar and the magnetosphere impart a torque \citep{cheng87},
to name two possibilities. 

In this paper, we present three coherent timing solutions for \psr\ spanning 6.4\,yr
with data obtained from the Green Bank Telescope (GBT), the Jodrell Bank
Observatory (JBO) and the \textit{Rossi X-ray Timing Explorer} (\rxte). 
We discuss two large glitches and timing noise found in these data.
We also present an analysis of the high energy
emission of the pulsar from the \rxte\ data by examining the X-ray pulse
profile from 2~--~40\,keV
and the phase offset between the radio and X-ray pulses.

\section{OBSERVATIONS AND ANALYSIS}
\label{sec:obs}

\begin{figure}
\plotone{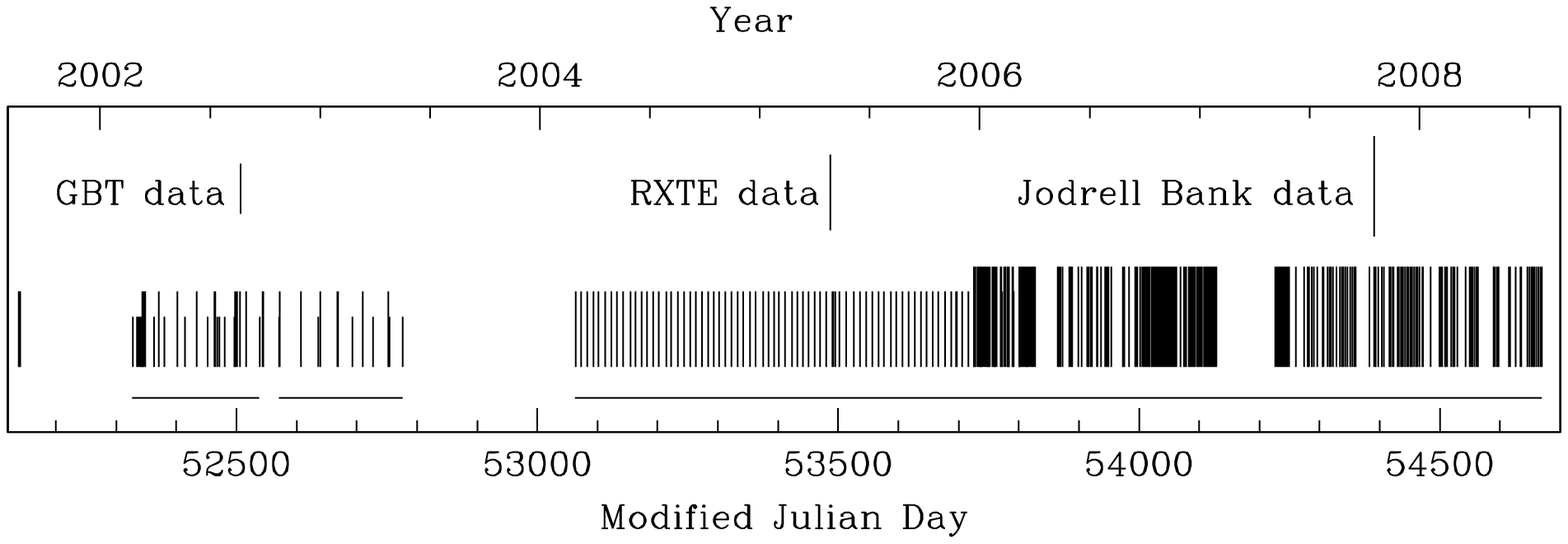}
\figcaption[dist]{Distribution of observations of \psr\ from \rxte, 
GBT, and JBO over 6.9\,yr. GBT data are indicated with short lines,
\rxte\ data with medium length lines, and JBO data with long lines. The 
three coherent timing solutions, spanning a total of 6.4\,yr,
are indicated with horizontal lines
along the bottom of the plot. 
\label{fig:dist}}
\end{figure}

X-ray observations were taken with \rxte; radio observations were taken with
the GBT and JBO, as detailed below. The data were unevenly
spaced throughout 6.9\,yr, as shown in Figure
\ref{fig:dist}, and include a 202-day gap (between observations 
taken in \rxte\ Cycle 6 and Cycle 7)
and a 287-day gap (corresponding to Cycle 8). GBT data are shown
with short lines, \rxte\ data are shown with medium-length lines,
and JBO data are shown with long lines. The first set of observations
from \rxte\ Cycle 6 are only considered for the pulse profile analysis
(\S\ref{sec:profile}) and
are not included in the 6.4\,yr of data for our 
timing analysis (\S\ref{sec:coherent}). 

\subsection{\rxte\ Observations and Analysis}
\label{sec:xteobs}

Observations of \psr\ were made using the Proportional Counter Array 
\citep[PCA;][]{jmr+06} on board \rxte. The PCA is an
array of five collimated 
Xenon/methane multi-anode proportional counter
units (PCUs) operating in the 2~--~60\,keV range, with a total effective
area of approximately $\rm{6500~cm^2}$ and a field of view of 
$\rm{\sim 1^o}$~FWHM.

Data were obtained during \rxte\ observing Cycles 7, 9, and 10, spanning a
period of 4\,yr from MJD 52327 to 53813, with a long gap (from MJD
52752 to 53036) corresponding to \rxte\ observing Cycle 8. In addition,
four observations were taken in observing Cycle 6 (17.1 hr exposure
over MJD~52138~--~52141). Because of the long data gap between the Cycle 6 data
and the subsequent observations in Cycle 7 (202 days), they could not be
unambiguously phase connected, rendering these observations of limited
interest for our timing analysis. They were considered for a pulse
profile analysis discussed in \S\ref{sec:profile}. 
The frequency for each Cycle 6 observation was
determined by a periodogram analysis, which was then used for folding the data.
All of the \rxte\ 
data were collected in ``GoodXenon'' mode, which records the arrival time
(with 1-$\mu$s resolution) and energy (256 channel resolution) of every
unrejected event. Typically, 3 PCUs were operational
during an observation. For our timing analysis, we used only the first layer of each operational
PCU in the energy range 2~--~18\,keV, as this maximizes the signal-to-noise ratio
of individual observations for this source. 

Observations were downloaded from the 
HEASARC archive\footnote{http://heasarc.gsfc.nasa.gov/docs/archive.html}
and photon arrival times were converted to barycentric dynamical time (TDB) 
at the solar system barycenter using the J2000 source position determined
using data from {\textit{CXO}}, RA
= $02^{\rm{h}}$~$05^{\rm{m}}$\,$37\fs92\pm0\fs02$, Dec
$=64\degrees$\,$49\arcmin$\,$42.8\arcsec\pm0.72\arcsec$ 
\citep{shm02} and the JPL DE200 solar system ephemeris
with the FITS tool `faxbary'.

We noted that two observations directly following
the leap second occurring on 2006 January 1 had incorrect clock
corrections, confirmed by the \xte\ team (C. Markwardt,
private communication). This was fixed by adding a 1\,s
time jump to each pulse time-of-arrival obtained from these observations.

\subsubsection{A New X-ray Timing Technique}
\label{sec:ML}

\begin{figure}
\plotone{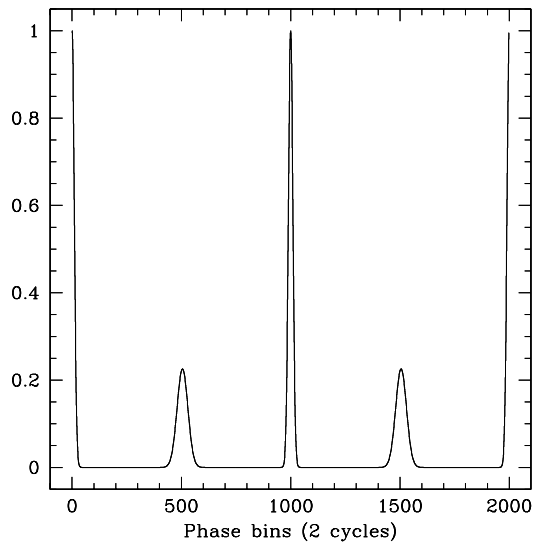}
\figcaption[overplot]{Best-fit two-Gaussian curve with 1000
phase bins from maximum-likelihood method of determining pulse times of arrival for
2~--~18\,keV as described in \S\ref{sec:ML}. 
\label{fig:gaussianprofile}}
\end{figure}

The X-ray pulse profile as measured in each observing session by \rxte\
has two narrow components separated by approximately one-half of a
rotation of the pulsar (See Fig.~\ref{fig:gaussianprofile}).  Such
sharp pulse features are
usually very helpful for timing analyses, however, the
``signal-to-noise'' ratio (here defined as the ratio of pulsed source
counts to all other counts) for the \rxte\ observations was very 
low, with typical values near 6$\times$10$^{-3}$.  These levels are so low
that for some observations the pulses were not easily visible
in the binned pulse profile plots.  Therefore, in order to 
determine the pulse times-of-arrival (TOAs) from the X-ray
data, we used a new and simple maximum-likelihood (ML) technique.
This method avoids the information loss inherent in binning
event-based (i.e. photon) pulse profiles.  The inputs to the method are an accurate
model pulse profile $I(\Phi)$ of intensity, including both signal and
the average noise level or background, as a function of rotational
phase $\Phi$ (where $0\le\Phi<1$), and the computed rotational phases
$\phi_i$ of the $N$ events (where $1\le i \le N$) from the observation
according to the best timing model of the pulsar.  We normalize
$I(\Phi)$ so that it has unit area and can be treated as a probability
density function for the individual event arrival times.

If we assume an arrival offset $t/P$ (where $P$ is the spin period)
of rotational phase (where $0\le
t/P<1$), we can compute a probability or likelihood that our data has
that offset from the template using ${\rm{Prob}}(t/P) = \prod_{i=1}^N
I(\phi_i-t/P)$.  If we compute a large grid of 1000-10000 evenly-spaced
probabilities for offsets $t/P$ between 0 and 1, the resulting
distribution describes the probability density for the average pulse
arrival time.  When normalized, we can determine the most likely
arrival time, or by suitably integrating the distribution, the median
arrival time and error estimates for the arrival time.

To test how well the ML technique determines TOAs compared to the
traditional binned Fourier-based methods typically used in radio
pulsar timing \citep{tay92}, we simulated
$\sim$100,000 high-energy observations of pulsars with both a pulse profile
like that of \psr\ (i.e. two narrow Gaussian peaks separated by nearly
half a rotation, Fig.~\ref{fig:gaussianprofile}) and a wide Gaussian profile with FWHM$=$0.2, each
with a wide variety of signal-to-noise ratios, total photon counts,
and numbers of bins in the profiles used to measure the pulse phase
for the traditional Fourier method.

The simulations showed that: 1. The traditional Fourier-based
technique underestimated the TOA errors by 10\% to several hundred
percent depending on the signal-to-noise ratio of the profiles as well
as the number of bins used in the pulse profile.  In contrast, the
average errors estimated by the ML technique were almost always within
5\%, and were typically within 1$-$2\%, of the true TOA errors as
determined from the statistics of the simulations.  2. The measured
TOA error distributions were narrower for the ML technique than for
the Fourier technique (i.e. the TOAs were more accurate) by several percent
typically, but by up to many tens of percent for low signal-to-noise
ratio cases where the ratio of pulsed counts to background counts is below
$\sim0.01-0.05$. 3. In very low signal-to-noise 
cases (i.e. where the ratio of pulsed counts to 
background counts is $\sim 0.001-0.01$), the Fourier method
determined completely wrong TOAs (i.e. where the measured TOA differed
from the true TOA by many times the estimated TOA error) much more
often than the ML technique, typically by factors of 2-5 times.

In summary, the ML technique determines slightly more accurate TOAs,
with much better error estimates, over a wider range of
signal-to-noise ratios, than the traditional binned Fourier technique
of TOA determination.  We recommend that it be used in all X-ray and
$\gamma$-ray pulsar timing applications, particularly in low
signal-to-noise ratio situations.

\subsubsection{Application to \psr}

For the ML-timing of \psr, we used a two-Gaussian profile model or
template $I(\Phi)$ based on a high signal-to-noise ratio pulse
profile from four months of \rxte\ data from early 2004 (Cycle 9). 
The timing model used for the profile included several frequency
derivatives such that no timing noise was apparent in the timing 
residuals. The two-Gaussian plus DC component model was then fit to
the data. The parameters of the resulting model are shown in 
Table~\ref{table:templateprofile}. The Gaussian template with 1000 phase
bins is shown in Figure~\ref{fig:gaussianprofile}.

For each \rxte\ observation, we folded the X-ray data using the best
predicted spin period for that day with the software package {\tt
  PRESTO} \citep{ran01}, but we allowed the software to search in a
  narrow range for the best pulsation period.  This technique was
  necessary to optimize the signal-to-noise ratio of the folded profiles
  because of the large levels of timing noise from the pulsar.  We then
  used the refined spin period to determine pulse phases $\phi_i$ for
  each of the X-rays.
  
  To determine the ML-derived TOAs, we used a grid of 1000 evenly spaced
  phase offsets $t/P$ over a full rotation of the pulsar.  While this is a
  fairly computationally expensive task when the number of events is
  large (as is the case for \rxte\ with its high background rate), the
  resulting likelihood distribution was typically well behaved (i.e.
  unimodal and nearly symmetric) and produced excellent TOAs and error
  estimates.  Typical uncertainties were between 450$-$750\,$\mu$s for
  each TOA.
  
  The phase offset, as determined by the median of the resulting
  likelihood distribution, was multiplied by the current pulse period
  and added to the fiducial epoch for each observation, in our case the
  first X-ray recorded, to make a TOA. Approximate one-sigma error
  estimates for each TOA were determined by integrating the likelihood
  distribution in each direction until 0.8413 of the total likelihood
  was accounted for.  For each \rxte\ observation, we typically determined
  2$-$3 TOAs.  The resulting TOAs were fitted to a timing model (see
  \S3.1 and \S3.2) using the pulsar timing software package
  TEMPO\footnote{http://www.atnf.csiro.au/research/pulsar/tempo/}.

\subsection{Green Bank Telescope Observations}

Observations with the GBT were made at either
820 or 1400\,MHz from MJD~52327~--~52776 using the
Berkeley-Caltech Pulsar Machine \citep[BCPM][]{bdz+97}.  The BCPM is an
analog/digital filterbank which samples each of 2$\times$96 channels
using 4 bits at flexible sampling rates and channel bandwidths.  
We recorded data using 134\,MHz of bandwidth and 50\,$\mu$s samples for
the 1400\,MHz observations or 48\,MHz of bandwidth and 72\,$\mu$s
samples for the 820\,MHz observations.  Typical integrations times
lasted between 3$-$5\,hrs. We de-dispersed at the known dispersion
measure (DM) of 140.7$\pm$0.3\,pc\,cm$^{-3}$ \citep{csl+02}
and folded all of the data using
{\tt PRESTO} \citep{ran01}, and then extracted 2$-$3 TOAs per
observation by correlating in the frequency domain the folded 64-bin pulse
profile with a Gaussian of fractional width 0.04 in phase.  The
typical precision of the TOAs was 200$-$400\,$\mu$s. 
TOAs were corrected to the UTC timescale using data from
Global Positioning System (GPS) satellites.

\subsection{Jodrell Bank Observatory Observations}

Observations were made at JBO every 3$-$5 days between MJD~53725 and
54666 using the  76-m Lovell telescope of the University of
Manchester at a frequency of 1.4\,GHz. 
Each observation typically lasted 3\,hr, divided into 1-minute
sub-integrations.  The data were de-dispersed at the known DM
in hardware and folded on-line. The profiles from these, sampled 
in intervals of 164.3\,$\mu$s, were added in polarization pairs and then combined to
provide a single total-intensity profile.  This was then convolved
with a template derived from a single high signal-to-noise-ratio
400-bin profile at the same frequency to yield a TOA.  TOAs were corrected to UTC
using information from the GPS. Further
details can be found in \citet{hlk+04}. The typical precision of the TOAs
was 200$-$700\,$\mu$s.

\section{TIMING ANALYSIS}

\subsection{Phase-Coherent Timing Analysis}
\label{sec:coherent}

The most accurate method used to determine spin parameters is phase-coherent
timing, that is comparing TOAs with a model ephemeris and accounting for
each rotation of the pulsar, as described elsewhere \citep[e.g.][]{ls05a}. 
A single phase-coherent timing solution spanning all 6.4\,yr of data proved
impossible for these data owing to the 287-day gap in timing observations as well
as two large glitches. As such, we present three coherent timing solutions,
summarized in Table~\ref{table:parameters}. 
We verified that changing the pulsar position by 3$\sigma$ did not
significantly change the fitted parameters, thus, the pulsar position was held 
fixed at that determined by \citet{shm02} from
{\textit{CXO}} data.

\begin{figure}
\plotone{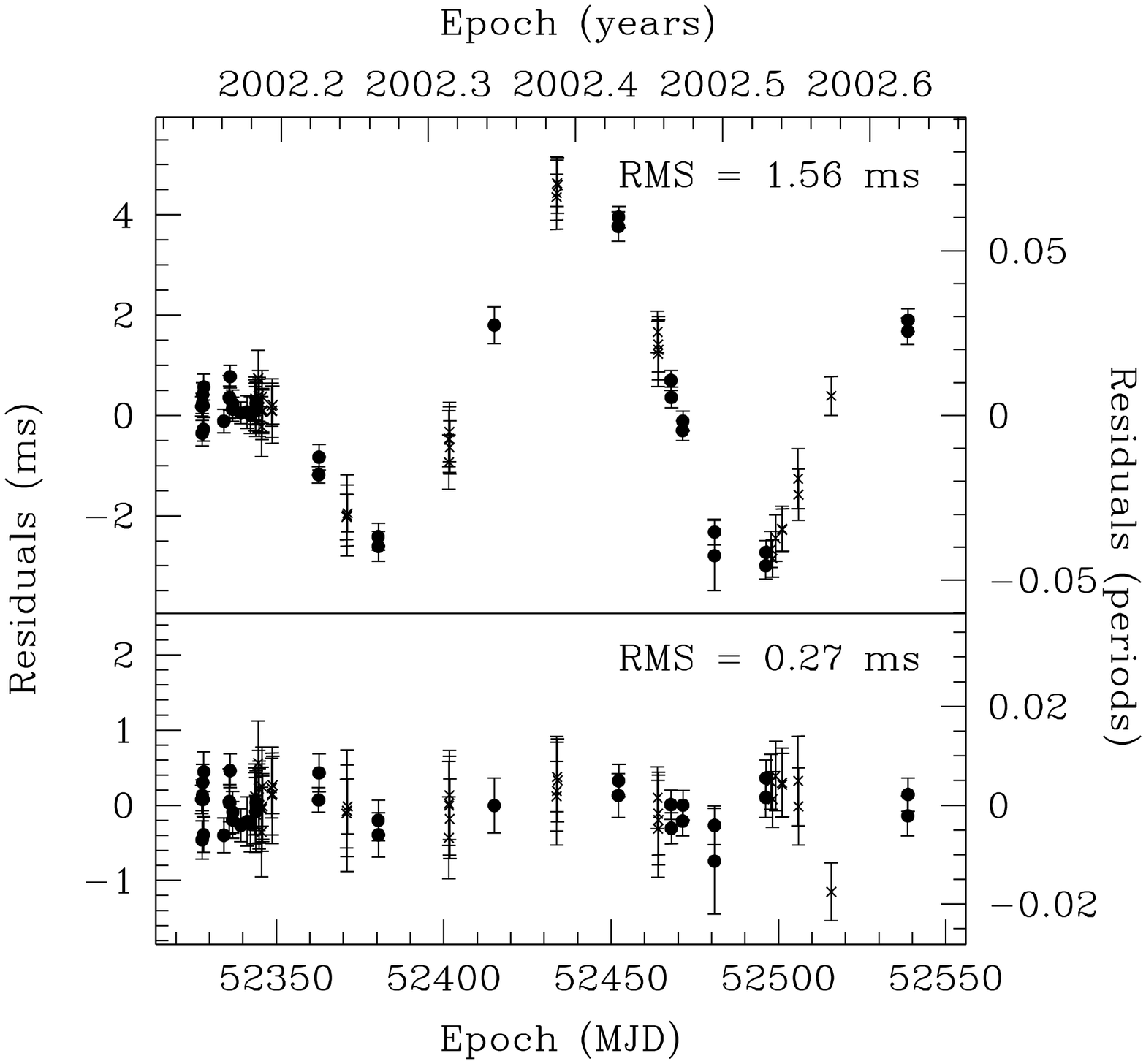}
\figcaption[c7_preglitch_resids.eps]{Timing residuals for \rxte\ observing
Cycle 7 (crosses) and GBT data (dots) spanning MJDs~52327~--~52538.
The top panel shows residuals with $\nu$, $\dot{\nu}$, and $\ddot{\nu}$
fitted. The bottom panel shows residuals with an additional four frequency
derivatives removed.  \label{fig:c7_preglitch}}
\end{figure}

GBT and \xte\ data were fitted together, resulting in a timing solution spanning six
months (MJD~52327~--~52538), with GBT (dots) and \xte\ (crosses) as shown in
Figure~\ref{fig:c7_preglitch}. Timing residuals are
shown in the Figure, with the 
top panel showing residuals with $\nu$, $\dot{\nu}$, and $\ddot{\nu}$ fitted. 
Significant timing noise remains in the data and can be fitted with six
frequency derivatives, shown in the bottom panel of
Figure~\ref{fig:c7_preglitch}. 

\begin{figure}
\plotone{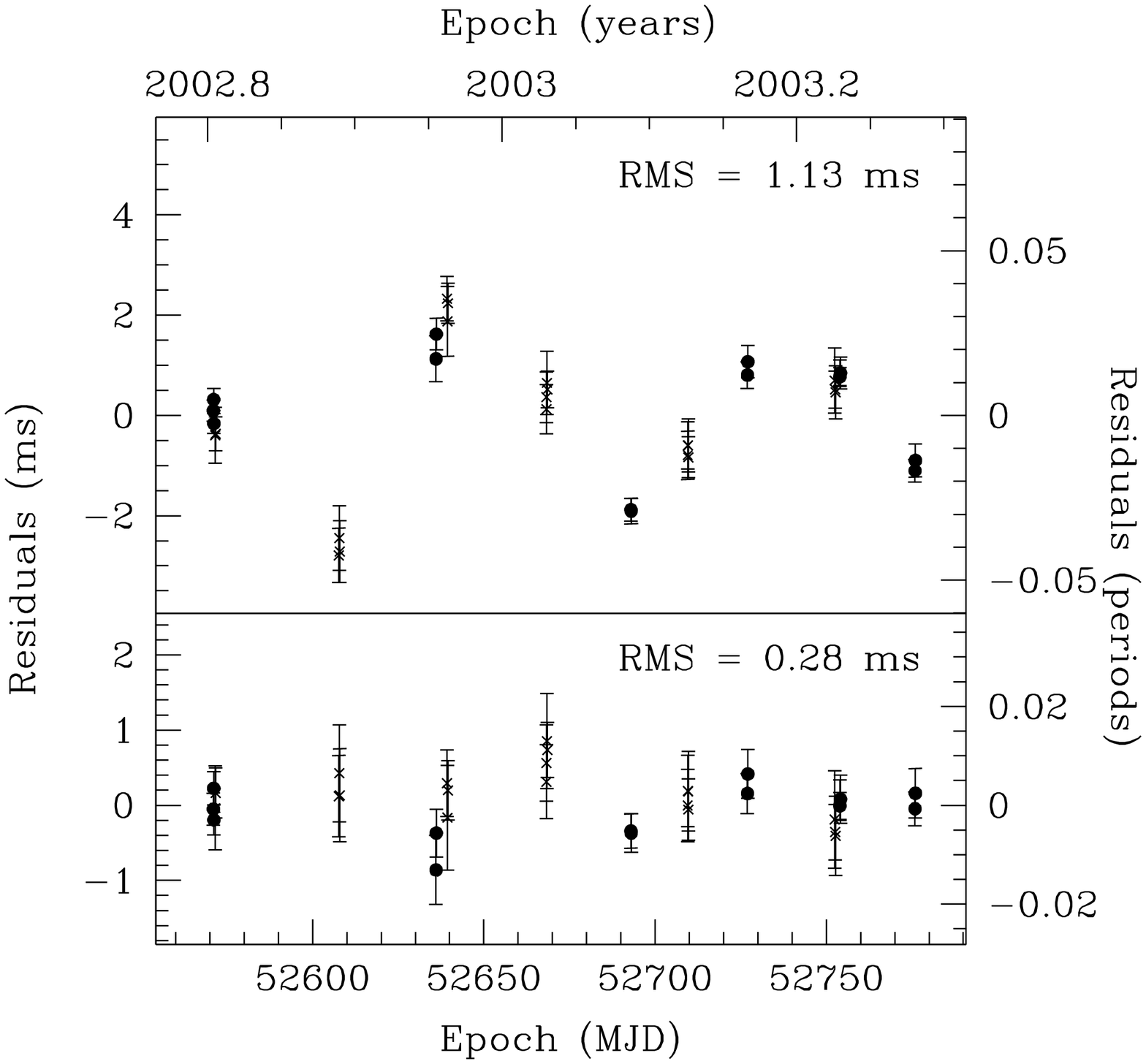}
\figcaption[c7_postglitch_resids.eps]{
Timing residuals for \rxte\ observing Cycle 7 (crosses) and GBT data (dots)
spanning MJDs~52571~--~52776. The top panel shows residuals
with $\nu$, $\dot{\nu}$, and $\ddot{\nu}$ fitted. The bottom panel shows
residuals with an additional three frequency derivatives removed.
\label{fig:c7_postglitch}}
\end{figure}

Phase coherence was lost after MJD~52538, as the result
of a glitch (see \S\ref{sec:glitches}). A second coherent 
timing solution using
GBT and \rxte\ data spans seven months (MJD~52571~--~52776). Timing residuals
are shown in Figure~\ref{fig:c7_postglitch}, with $\nu$, $\dot{\nu}$, and $\ddot{\nu}$ fitted
in the top panel. As with the previous timing solution, significant 
timing noise remains in the residuals, which is fitted with 5 frequency
derivatives, shown in the bottom panel of Figure~\ref{fig:c7_postglitch}.

\begin{figure}
\plotone{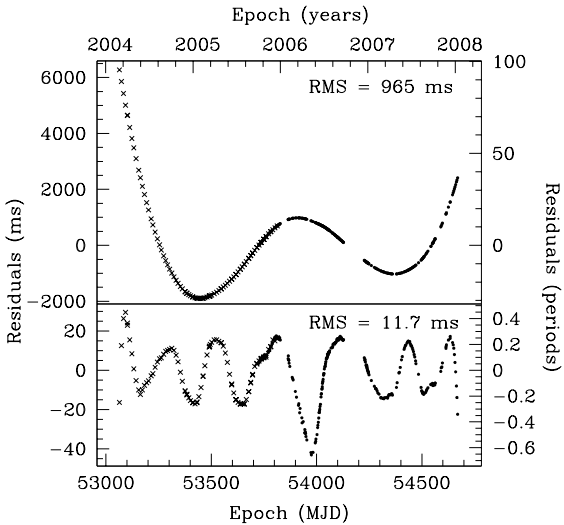}
\figcaption[bw_xte_jb_resids.eps]{Timing residuals for \rxte\ observing
Cycles 9 and 10 (crosses),
and Jodrell Bank data (small dots), after the glitch occurring between
MJD~52777 and 53062.
The top panel shows residuals with only $\nu$, $\dot{\nu}$, and $\ddot{\nu}$
fitted. The bottom panel shows residuals with $\nu$ and 12 frequency
derivatives
fitted (the maximum allowed with current machine precision).
\label{fig:cycle9_10_JB}}
\end{figure}

Figure~\ref{fig:cycle9_10_JB} shows residuals for the third timing
solution spanning $\sim$4.4\,yr (MJD~53063~--~54669). 
In addition to two years of X-ray timing observations (spanning MJD~53063~--~53813),
on MJD~52725 (2005 December 21), radio timing observations began using JBO.
X-ray and radio observations were concurrent for 88 days, after which no additional
X-ray observations were obtained. 

The top panel of Figure~\ref{fig:cycle9_10_JB} shows residuals with 
$\nu$, $\dot{\nu}$, and $\ddot{\nu}$ removed; the bottom panel shows
residuals with 12 frequency derivatives fitted. The timing noise 
in this 4.4-yr period
is so large that it cannot be fully described by a 12 degree polynomial
(the largest allowed with current machine precision). In addition,
some of the timing noise seen in
these data is likely attributable to unmodeled glitch recovery
\citep{lyn96}. 
In fact, the measured value of  $\nu$ 
is significantly different from that predicted from the previous timing solution. 
The difference between the predicted and measured $\nu$ is too large
($\sim$6$\times10^{-5}$\,Hz) to be explained by 
timing noise, indicating that a glitch probably occurred during the 287-day gap 
in the data. 

\subsection{Glitches}
\label{sec:glitches}
In general,
the frequency evolution of a pulsar following a glitch can be characterized by
\begin{equation}
 \nu(t) = {\nu}_0(t) + \Delta{\nu} [ 1 - Q(1 - e^{-t/{\tau_d}})] +
\Delta{\dot{\nu}_p}t,
 \label{eqn:glitch}
 \end{equation}
where ${\nu}_0(t)$ is extrapolated from the pre-glitch value,
$\Delta{\nu}$ is the initial jump in frequency,
$Q$ is the recovery fraction, $\tau_d$ is the time decay constant,
and $\Delta{\dot{\nu}_p}$ is
the permanent change in $\dot{\nu}$ as a result of the glitch.

\begin{figure}
\plotone{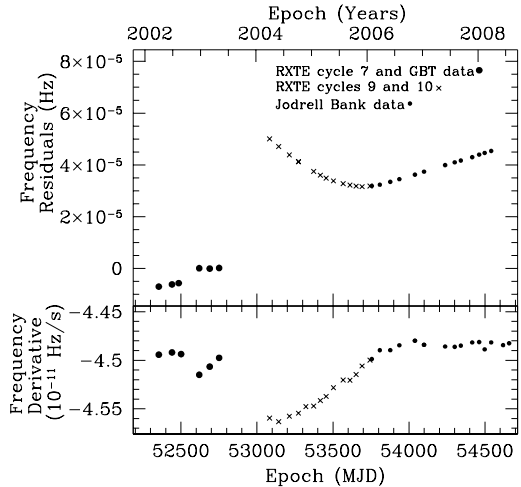}
\figcaption[3c58_nu_nudot_evolution.eps]{Frequency and frequency derivative
evolution of \psr\ over 6.4\,yr. The top panel shows frequency measurements from short,
phase-coherent timing solutions with the overall trend in the inter-glitch
interval subtracted. The bottom panel shows measurements of the frequency derivative.
Uncertainties are smaller than
the plotted points. \label{fig:partial_nu}}
\end{figure}

In order to analyze further the two glitches inferred from our coherent
timing analysis, we performed a partially coherent timing analysis 
over short time intervals (on average $\sim$40\,days), fitting 
only for $\nu$ and $\dot{\nu}$ and
choosing the length of each data subset such that the phase residuals are
Gaussian-distributed (i.e. `white'). The results, with
the average $\dot{\nu}$ from the inter-glitch period
removed, are shown plotted in the top panel of Figure~\ref{fig:partial_nu}.
Two glitches as well as timing noise are clearly present. We also show
measurements of $\dot{\nu}$ for the same intervals 
in the bottom panel of Figure~\ref{fig:partial_nu}. Again, two
glitches are apparent, as is the significant timing noise in the data. 

\begin{figure}
\plotone{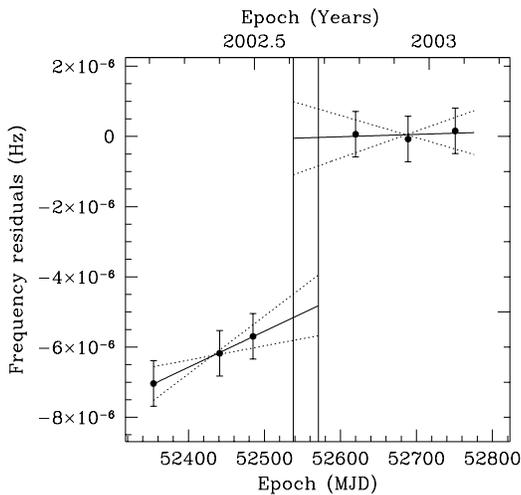}
\figcaption[cycle7_glitch.eps]{Frequency measurements with the post-glitch
trend subtracted, showing the first observed glitch occurring between MJDs~52538 and
52571 (indicated by vertical lines) with fractional magnitude of
$\Delta{\nu}/{\nu} =(3.4\pm1.1)\times10^{-7}$. $1~\sigma$ uncertainties
in the pre- and post-glitch slopes are shown with hatched lines. The change 
in $\dot\nu$ as measured from the difference in the slope before and
after the glitch is not statistically significant. 
\label{fig:cycle7_glitch}}
\end{figure}

\begin{figure}
\plotone{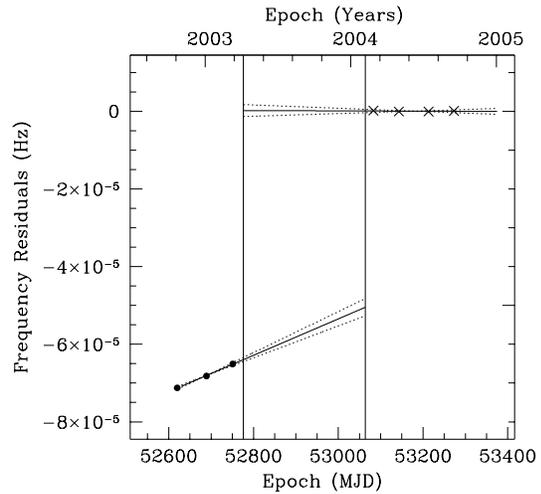}
\figcaption[cycle8_glitch.eps]{Frequency measurements with the post-glitch
trend subtracted, showing the frequency increase occurring between
MJDs~52777 and
53062 (indicated by vertical lines). This could be the result of two or more
separate glitches occurring during 287-day gap in the data, however, we
interpret the frequency jump as the likely result of a single
glitch of fractional magnitude $\Delta{\nu}/{\nu} =(3.8\pm0.4)\times 10^{-6}$.
One-$\sigma$ uncertainties on pre-
and post-glitch slopes are shown with hatched lines and are indicative of a
change in ${\Delta{\dot{\nu}}}/{\dot{\nu}} = 0.012\pm0.001$.
\label{fig:cycle8_glitch}}
\end{figure}

To measure accurately the size of each glitch while minimizing the
contaminating effect of long-term timing noise, we took $\nu$
measurements spanning only $\sim$200 days before and after
each glitch to measure the 
fractional increase in spin frequency. Figures~\ref{fig:cycle7_glitch} and
\ref{fig:cycle8_glitch}
show pre- and post-glitch $\nu$ measurements with the post-glitch slope
subtracted. 

We observed a frequency increase between MJDs~52538 and 52571
(Fig.~\ref{fig:cycle7_glitch}), corresponding to the loss of phase
coherence discussed in \S\ref{sec:coherent}.
This $\nu$ increase corresponds to a glitch of fractional magnitude
$\Delta{\nu}/{\nu} = (3.4\pm1.1)\times10^{-7}$ (see also 
Ransom et al. 2004\nocite{rck+04}). The 
change in ${\dot\nu}$ over the glitch is not statistically 
significant when measured as
the slope of frequency measurements, as shown in the Figure.
However, the fractional change in $\dot{\nu}$ from the individual
$\dot{\nu}$ measurements before and after the glitch is significant, with
$\Delta{\dot{\nu}}/{\dot{\nu}}\sim0.005\pm0.001$. 
No short-term post-glitch relaxation is detected; however, because of the sparse
sampling, it cannot be precluded. Neither can a long-term post-glitch
relaxation be distinguished from a simple change in $\dot\nu$ because
a decay, if present, was interrupted by a second glitch. 

We observed a second frequency increase between MJDs~52776 and 53063 as shown in
Figure~\ref{fig:cycle8_glitch}.  Because no timing data were taken during
the 287-day period where the frequency jump occurred, it is not possible
to differentiate between a single glitch and two (or more) smaller glitches.
However, because there is no clear evidence of more than one glitch, we
interpret the frequency increase as a single glitch of fractional magnitude
$\Delta\nu/\nu = (3.8\pm0.4) \times 10^{-6}$. The frequency derivative, as 
measured from the slope before and after the glitch (see
Fig.~\ref{fig:cycle8_glitch}) also changed significantly, with a fractional magnitude of 
$\Delta{\dot{\nu}}/{\dot{\nu}}=0.012\pm0.001$. 

\begin{figure}
\plotone{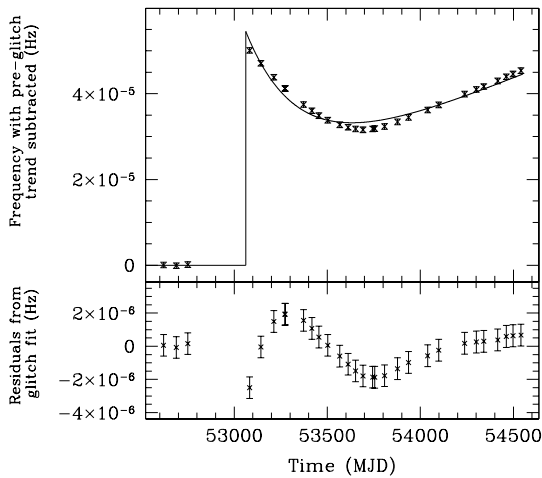}
\figcaption[glitchresids.eps]{Pulse frequency measurements 
and fitted glitch model for the large
glitch occurring between MJDs~52777 and 53062 for \psr.
The top panel shows frequency measurements with pre-glitch trend
removed, with the fitted model for a typical glitch epoch 
over-plotted, while the bottom panel shows the
residuals. The uncertainties on each point
are dominated by the uncertainty in the pre-glitch trend subtracted
from the data and therefore are of roughly uniform size.
For this fit to the data, the glitch
epoch is MJD~53062, with $\Delta{\nu}/\nu\sim 3.6\times10^{-6}$,
$Q\sim0.66$, 
$\tau_d\sim280$ and $\Delta{\dot{\nu}}/{\dot{\nu}} \sim-0.00453$, resulting
in ${{\chi}^2}_{\nu} = 3.9$ for 25 degrees of freedom.
\label{fig:glitchresids}}
\end{figure}

Another possible description of this glitch is a large increase in $\nu$
followed by an exponential recovery (also suggested for this glitch by
W. Hermsen, personal communication). Fitting a glitch model including an 
exponential decay to all the frequency measurements before and
after the glitch, we found models which span a
wide range of possible glitch parameters, with
a reduced ${\chi^2}$ of $\sim 3.6-3.9$ for 25 degrees of freedom. 
We found fits of roughly equal probability for all possible glitch
epochs between the two bounding coherent
timing solutions, thus we present a range of glitch parameters
corresponding to the date limits of MJD 52777 and MJD 53062. 
A typical fit and residuals are shown in Figure~\ref{fig:glitchresids}. 
The fractional magnitude of the glitch from these models ranges from
$\Delta{\nu}/{\nu}\sim (3.6 - 7.2)\times10^{-6}$, while the 
recovery fraction spans $Q\sim 0.66-0.88$ and $\tau_d \sim 280 - 295$. 
The long-term change in
frequency derivative, $\Delta{\dot{\nu}}/{\dot{\nu}}$, is always negative,
i.e. in the opposite direction from that expected in a typical glitch,
and is approximately equal to $\sim$$-0.0046$ for all fitted models. 
This could be attributed to unusual glitch
recovery, timing noise, or a combination thereof. However, this effect is
possibly an artifact of the fitting procedure, where it is assumed that the
pre-glitch $\nu$ is not itself recovering from the previous glitch.

\begin{figure}
\plotone{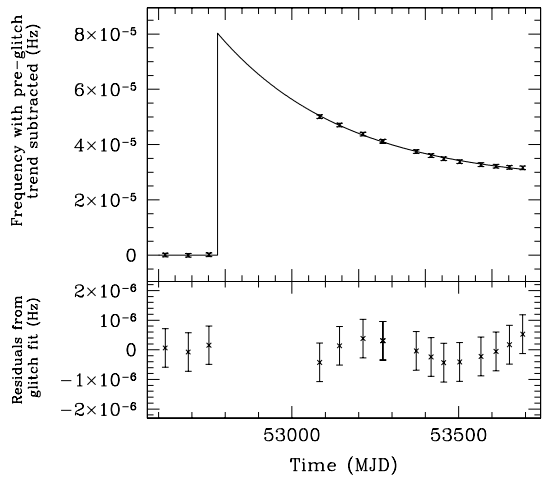}
\figcaption[shortglitchresids.eps]{Pulse frequency measurements and
fitted glitch model for a typical glitch epoch for \psr,
considering only the period of glitch recovery and ignoring the long-term
post-glitch change in $\dot\nu$. The top panel shows frequency
measurements with the pre-glitch trend removed, with the fitted model
over-plotted, while the bottom panel shows the residuals from the fit. 
The uncertainties on each point
are dominated by the uncertainty in the pre-glitch trend subtracted
from the data and therefore are of roughly uniform size.
Shown here is the best fit model for a glitch occurring 
on MJD 52777 with $\Delta\nu/\nu
\sim 5.3\times 10^{-6}$, $\tau_d \sim 383$, and $Q=0.67$. The reduced
$\chi^2$ for the fit is $0.262$ for $12$ degrees of freedom. 
\label{fig:shortglitch}}
\end{figure}

Because of the unusual behavior of the long-term $\Delta\dot\nu$ after the
glitch, and the possibility that this behavior is not a direct result
of the glitch, we performed a third fit to the frequency data, this
time excluding frequency measurements after MJD 53700 (after which the
unusual $\Delta\dot\nu$ dominates), and fixing the
change in $\dot\nu$ to be zero. Fitting an exponential glitch recovery
model to this subset of data provides a much better fit,
as shown for a sample glitch epoch in Figure~\ref{fig:shortglitch}. 
Again, the glitch may have
occurred at any time between MJD 52777 and 53062. For this fit, the
range of possible fractional increases in frequency is $\Delta\nu/\nu =
(3.4-5.3)\times10^{-6}$, while the recovery fraction is $Q= 0.5-0.67$
and the recovery time scale is $\tau_d = 379-383$\,days.

\section{X-RAY PROFILE ANALYSIS}
\label{sec:profile}
For each \xte\ observation in observing Cycles 7, 9, and 10, we created
a phase-resolved spectrum with 64 phase bins across the profile, using the Ftool
`fasebin' and the partially coherent timing ephemerides described above
(\S\ref{sec:glitches}). 
We also created phase-resolved spectra as described above
for the four observations taken
during \rxte\ Cycle 6. These observations could not be unambiguously
phase-connected with the later data because of a 202-day gap between
these and the subsequent Cycle 7 
observations. Instead, we folded the observations at a frequency
determined by a periodogram analysis, $\nu=15.22466(2)$\,Hz with a frequency
derivative of $\dot\nu=-4.49(1)\times10^{-11}$\,s$^{-2}$ evaluated at
the reference epoch
MJD~52139.5. Because of the hard spectrum of the
source \citep{rck+04}, we selected photons from all three layers of the
Xenon detectors, and from all operational PCUs and included all photons
within the energy range 2~--~60\,keV. In addition, we repeated the entire
process using the first Xenon layer for the energy range 2~--~18\,keV, where the
majority of the softer source counts reside.
We also created phase-averaged spectra which were used 
to build response matrices for each PCU. We then used XSPEC\footnote{See
http://xspec.gsfc.nasa.gov, ver. 11.3.1.} to 
create a pulse profile in count rate for each PCU for each 
observation. We then combined data from all PCUs for each observation,
dividing by a factor to account for the amount of time each PCU was on,
resulting in a pulse profile in units of count rate per PCU.

\begin{figure}
\plotone{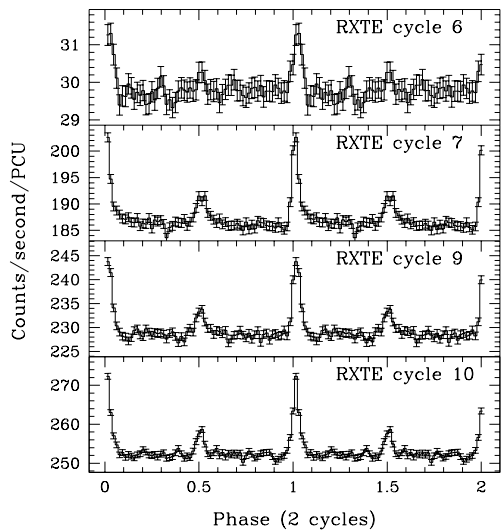}
\figcaption[histogram_time_evolution.eps]{
Pulse profiles of \psr\ 
for \rxte\ Cycles 6, 7, 9 and 10 for the energy range 2~--~18\,keV.
Note that \rxte\ Cycle 6 comprised significantly less
observing time, resulting in larger uncertainties. The interpulse is not
clearly visible in the Cycle 6 data, but a $\chi^2$ test shows that the
profile is not significantly different from the Cycle 7, 9, or 10 pulse
profiles. 
\label{fig:profevolution}}
\end{figure}

We aligned the 2~--~18\,keV profile for each observation
by cross-correlating
with the 2-Gaussian template (Fig.~\ref{fig:gaussianprofile})
and summed profiles from each \rxte\ cycle together, as shown
in Figure~\ref{fig:profevolution}. There is no evidence for 
evolution in the profile over $\sim$4\,yr.

\begin{figure}
\plotone{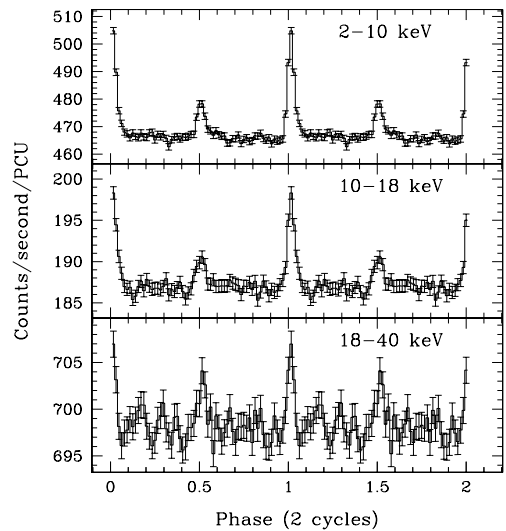}
\figcaption[hist_energy_evolution.eps]{
Pulse profile of \psr\ shown in energy bands 2~--~10\,keV,
10~--~18\,keV, and 18~--~40\,keV. Each pulse profile is created by
aligning and summing individual profiles from \rxte\ Cycles 7, 9, and
10. Visible structure in the off-pulse region of the 
18~--~40\,keV profile is not statistically significant as shown by a
$\chi^2$ test, a $Z^2$ test, and a $H$ test. 
\label{fig:energyevolution}}
\end{figure}

We also created profiles for different
energy bands: 2~--~10\,keV, 10~--~18\,keV, and 18~--~40\,keV and summed
profiles from \xte\ cycles 7, 9, and 10 (where phase-coherent timing
solutions exist) into a single profile for each energy band,
shown in Figure~\ref{fig:energyevolution}. We used the phase offset
determined from cross correlating the 2~--~18\,keV profile to align
the profiles from all the energy
ranges. This ensures that the correct phase offset is applied the
18~--~40\,keV profiles, where the signal-to-noise ratio of individual
profiles is very poor, and a noise spike can be mistaken by the cross
correlation algorithm for the sharp peak of the profile. We note that
using the phase offset determined for a different energy range profile
could be problematic in building added profiles if there is signifcant
energy evolution across the energy range of interest. This does not
appear to be the case for this pulsar for the relevant energy range.
The pulsar is visible up to
$\sim40$\,keV and the pulse shape does not vary significantly with
increasing energy. 

There is some possible unusual structure in the off-pulse region of the
profiles, particularly for the energy range 18~--~40\,keV.
We examined this putative structure by performing several additional analyses. 
First, we produced summed profiles using a different method than 
described above. We extracted events and created a time series
for each energy range, for each individual observation. We folded
each time series with the local ephemeris and aligned and summed
the profiles several times using different template profiles. This
analysis produced pulse profiles that are not significantly 
different from those shown in Figure~\ref{fig:energyevolution}.

We further analyzed the off-pulse region of the profiles by applying
a $\chi^2$ test, a $Z^2$ test \citep{bbb+83}, and an $H$ test
\citep{jrs89}. The reduced $\chi^2$ values for the
10~--~18\,keV and 18~--~40\,keV are less than 1, while the 
$\chi^2$ for the off-pulse region of the 2~--~10\,keV profile is 55
for 48 degrees of freedom (where the probability of this $\chi^2$
or higer occurring by chance is 21\%). The $Z^2$ test for 1, 2, 4, and 8 
harmonics, and the $H$ test applied
to the off-pulse region of all three profiles resulted in the null
hypothesis. Therefore, the off-pulse regions of the profiles are not
statistically different from a DC offset. 

\section{PHASE OFFSET BETWEEN THE RADIO AND X-RAY PULSES}
Precise measurement of the phase lag between the radio and X-ray pulses
is important for understanding the pulse emission mechanism.
Emission from rotation-powered pulsars is thought to arise from 
either a polar cap \citep[e.g.][]{dh82}, 
or in magnetospheric outer gaps \citep[e.g.][]{chr86b,rom96a}. 
Absolute timing for several different energy ranges can
place constraints on the shape of the outer gap and the height 
in the magnetosphere where radiation
is generated \citep{ry95}. 
The radio pulse profile of \psr\ is single peaked with a width of $\sim$$0.05P$
\citep{csl+02}; the X-ray profile is double peaked, with a peak-to-peak
separation of $\sim$0.5. We made two independent measurements of the phase 
offset between
the radio pulse and the main X-ray pulse for \psr, by finding the offset 
between the GBT
and \xte\ data, and the JBO and \xte\ data.

We used the first timing solution (GBT and \rxte\ data prior to glitch 1)
spanning MJDs~52327 to 52538 to make three independent
measurements of the phase offset. We split our timing solution into three
subsets, fitting only for $\nu$ and $\dot{\nu}$. Radio TOAs were shifted to
infinite frequency using the nominal DM \citep{csl+02}.
The data subsets were chosen such that there was good overlap between
the sparsely sampled X-ray and radio data, and that each solution
had Gaussian-distributed residuals. The weighted average value of these three
measurements is $6.72\pm0.66$\,ms, or $0.102\pm0.010$ in phase (radio leading).
This analysis could not be repeated for the post-glitch GBT/\rxte\ timing solution
because of the even more sparsely sampled data, which consist of significant 
lags between most radio and X-ray observations, leading to poorly
constrained values of the phase offset for short sections of the data.

We obtained a second measurement of the phase offset by simultaneously
fitting overlapping timing observations from \xte\ and JBO, spanning 88 days
from MJDs~53725 to 53813. We first obtained a phase-coherent timing solution
from the well sampled radio data, and then added the overlapping, more
sparsely sampled X-ray TOAs. Again, radio TOAs were shifted to infinite
frequency using the nominal DM value. We split the data into three subsets, fitting
only for $\nu$ and $\dot{\nu}$ and ensuring that each subset had Gaussian
residuals. The JBO TOAs were created using a different 
fiducial point than the GBT and \rxte\ profiles, so each TOA was shifted by
a constant to adjust for the difference between the two fiducial points used. 
The weighted average of these offset measurements is
$5.55\pm0.66$\,ms, corresponding to a phase offset of $0.085\pm0.010$.
The difference between this measurement and that
made with \rxte\ and GBT data is $1.17\pm0.94$\,ms, i.e., in agreement within
$1.25\sigma$. 

We further verified the phase offset by extracting a single TOA from an 
archival {\textit{CXO}} observation from February 2002 (Observation ID 2756).
The offset measured from the {\textit{CXO}} TOA to the GBT radio TOAs
agrees with our GBT/\rxte\ offset within 1.2\,$\sigma$. 

The uncertainty in the phase offset is dominated by the uncertainty in the DM.
The DM was measured to be 140.7$\pm$0.3\,pc\,cm$^{-3}$ 
using 800 and 1375\,MHz
GBT data in 2002, as reported in \citet{csl+02}. Thus the JBO determination 
of the phase lag, made $\sim$3.5\,yr after 
the measurement of DM, could be affected by short or long-term changes in DM.
We therefore report the first measurement of the phase offset made with GBT and
\rxte, of $\phi=0.102\pm0.010$.

\section{DISCUSSION}
\label{sec:discussion}

\subsection{Timing Noise, Glitches, and the Age of \psr}
We show evidence for two large glitches in 6.4\,yr of timing data of \psr.
The fractional magnitudes of these glitches ($\Delta{\nu}/{\nu}\sim10^{-7}-10^{-6}$) are
typical of pulsars with characteristic ages of 5$-$10\,kyr, such as
the frequent large glitcher PSR J0537$-$6910 ($\tau_c=4.9$\,kyr) or the Vela
pulsar ($\tau_c=11$\,kyr). The youngest pulsars such as the Crab pulsar
(955\,yr), PSR B0540$-$69 ($\tau_c=1.7$\,kyr) and PSR J1119$-$6127
($\tau_c=1.6$\,kyr) typically have glitches with smaller fractional magnitudes
ranging from $\Delta{\nu}/{\nu}\sim10^{-9}-10^{-8}$.
Perhaps this indicates that the pulsar age is
closer to its characteristic age of $\tau_c \sim5.4$\,kyr, rather than the
historical supernova age of 828\,yr. 

If the pulsar was born in the historical supernova event 828\,yr
ago, these glitches are unusually large. Glitches may be related to pulsar
age via neutron star temperature \citep{ml90}. If this is the case,
the large glitches observed here could be related to the very low 
measured temperature of
\psr\ \citep{shsm04}, rather than its chronological age. 
The reason for the exceptionally cool 
surface temperature of this neutron star is still a mystery, though
may be explained with a large neutron star mass \citep{ykhg02}.
On the other hand, large glitches have been observed in the hot 
surface temperature Anomalous X-ray
Pulsars \citep[AXPs, e.g.][]{klc00,dkg07}. If the mechanism behind
rotation-powered pulsar and magnetar glitches is similar, the neutron star surface 
temperature may not be the primary factor in determining the size of glitches.

Young pulsars emit large amounts of energy as they spin down, 
providing for easy measurement of $\dot{\nu}$, and occasionally higher order
frequency derivatives
\citep[$\ddot{\nu}$, $\nudotdotdot$; e.g.][]{lps93,lkgm05}.
A measurement of $\ddot{\nu}$ provides an estimate of
the `braking' index, $n={\nu}{\ddot{\nu}}/{\dot{\nu}^2}$, 
giving insight into the physics underlying pulsar
spin-down, as well as an improved estimate of the pulsar age. Both timing
noise and glitches can prevent a measurement of $n$.
Generally, only the youngest pulsars, with $\tau_c<2$\,kyr,
have measurable braking indices. The exception is the 11\,kyr-old Vela
pulsar, where frequent large glitches prevent a phase-coherent measurement
of the braking index, but measuring $\dot{\nu}$ in the aftermath of glitches
has allowed a measurement of $n=1.4\pm0.2$ over $\sim$25\,yrs of 
data \citep{lpgc96}.

The initial goal of timing \psr\ was to measure its braking index. 
However, the measured value of $n$ varies significantly
among the three phase-coherent timing solutions obtained for this source,
ranging from $n\simeq 15-90$. A partially coherent timing analysis (Fig.~\ref{fig:partial_nu})
shows that $\dot{\nu}$ does not evolve linearly
implying that a deterministic value of $\ddot{\nu}$ and thus $n$ cannot
be measured from these data. However, excluding data in the immediate
aftermath of the glitches and looking at the overall trend in $\dot{\nu}$
from the first three and last 10 measurements of $\dot{\nu}$ from
Figure~\ref{fig:partial_nu} (bottom panel), the implied value is $n\sim4$.
Though this value is contaminated by timing noise
and glitch recovery, it is suggestive that the true, underlying value of 
$n$ may eventually be measurable with long-term timing.

\subsection{Absolute Timing and the Pulse Emission Mechanism}

The observed phase difference between radio and high-energy pulses, as well 
as the pulse shape and peak-to-peak separation, in principal provide important
information about the pulsar emission mechanism by constraining the pulse emission
region. Table~\ref{offset_table} shows the offsets
between the radio and X-ray pulses and the radio and $\gamma$-ray pulses for
all known measurements to date. These compiled data show that there
is a large scatter in the measured offsets and no
correlation between pulse period and phase offset. 

The measured offset for \psr\ of
$\phi=0.10\pm0.01$ supports the lack of correlation between phase
offset and pulse period. In particular, \psr\ has a very similar pulse
period (65.7\,ms) to the pulsar PSR~J1420$-$6048 (68.2\,ms), while the
phase offset of PSR~J1420$-$6048 is $\phi=-0.35(6)$ \citep{rrj01}. 
It is likely that the geometry and viewing angle of each system will 
affect the measured phase offset and may ultimately
explain the range of observed offsets. 

The radio-to-X-ray phase offset for \psr\ is consistent with
the radio-to-$\gamma$-ray offset of $\phi=0.08\pm0.02$ \citep{fermi_0205}.
The X-ray and $\gamma$-ray phase offsets are likewise aligned for the
Crab \citep{ppp+08} and Vela pulsars \citep[e.g.][]{fermi_vela}, while 
this is not the case for the young pulsar
PSR~B1509$-$58 \citep{khk+99}, nor the millisecond pulsar 
PSR~J0218+4232 \citep{fermi_msp}. 

The outer gap model of \citet{ry95} predicts that the X-ray pulse should lag
the radio pulse by 0.35$-$0.5 in phase and should appear as a single broad
pulse. Neither prediction is supported by the X-ray profile of \psr.
In addition, a thermal X-ray pulse arriving in
phase with the radio pulse is predicted, while no thermal pulsations have
been detected in \psr\ \citep{mss+02}. 

The main pulse of \psr\ is detected up to 40\,keV with the PCA on board \rxte, 
while the
interpulse is visible up to $\sim$18\,keV. It is one of an increasing number
of young pulsars detected in the hard X-ray energy range and
has also recently been detected up to $\sim3$\,GeV with the \textit{Fermi Space
Telescope} \citep{fermi_0205}. Comparing the emission of the Crab pulsar and
nebula to the 3C 58 pulsar and nebula at higher energies may offer insight
into the physical reasons behind the intriguing differences between these
two seemingly similar objects. 

\section{CONCLUSIONS}
Multi-wavelength timing observations offer an excellent
probe of both the temporal and emission characteristics of young pulsars. 
We observed two large glitches that are not characteristic of the proposed
young age of \psr. This is not conclusive however, and the timing data are
consistent with an age of 828\,yr if the pulsar was born spinning slowly 
with $P_0\sim60$\,ms, or an age of $\tau_c\sim5.4$\,kyr if the pulsar was born
spinning rapidly. Furthermore, the age of the pulsar is consistent with 
$\tau>\tau_c=5.4$\,kyr if its true braking index $n<3$, as is the case
for all
measured values of $n$, though there is currently no evidence for $n<3$ for
\psr. Long-term timing may eventually allow
for the measurement of $n$ by using the incoherent method performed on the Vela pulsar
\citep{lpgc96}.

We have presented the first measurement of the phase offset between the
radio and X-ray pulses of \psr\ to be $0.10\pm0.01$, which is
consistent with the recently reported 
$\gamma$-ray phase offset of $\phi= 0.08\pm0.02 $ \cite{fermi_0205}.
\psr\ is rare in that it has both the magnetospheric X-ray and
$\gamma$-ray phase offset precisely measured. Among the pulsars 
with measured phase offsets, with periods ranging
from 1.56\,ms to 1250\,ms, there is no correlation between pulse
period and phase offset, as shown in Table~\ref{offset_table}.
Phase offset measurements should be important for constraining the pulse emission region
and significant progress is occurring at present in the measurement of
radio-to-$\gamma$-ray phase offsets with the ongoing detection of many radio
pulsars as $\gamma$-ray pulsars with \textit{Fermi}.

\acknowledgments
We would like to thank an anonymous referee for helpful comments and
suggestions that greatly impoved our manuscript. 
This research made use of data obtained from the High Energy Astrophysics
Science Archive Research Center Online Service, provided by the NASA-Goddard
Space Flight Center. The National Radio Astronomy Observatory is a facility
of the National Science Foundation operated under cooperative agreement by
Associated Universities, Inc. MAL is an NSERC PGS-D fellow. VMK is a Canada Research
Chair and the Lorne Trottier Chair. 
Funding for this work was provided by 
NSERC Discovery Grant Rgpin 228738-03.  Additional funding came from 
Fonds de Recherche de la Nature et des Technologies du Qu\'ebec and
the Canadian Institute for Advanced Research. Pulsar research at UBC is
supported by an NSERC Discovery Grant. IHS acknowledges support from the
ATNF Distinguished Visitor program and from the Swinburne University
of Technology Visiting Distinguished Researcher Scheme.

\clearpage

\begin{center}
\begin{deluxetable}{ll}
\tablecaption{X-ray Profile Template Parameters \label{table:templateprofile}}
\tablewidth{0pt}
\startdata
Parameter & Value\\
\hline \hline
    Flux 1 &    0.64515\\
    FWHM 1      & 0.02386\\
    Phase 1     & 0.0\\
    Flux 2      & 0.35485\\
    FWHM 2      & 0.05826\\
    Phase 2     & 0.50515\\
    DC flux     & 193.46 \\
\hline
\enddata
\\
{Parameters for the two-Gaussian template profile, which is
proportional to the X-ray arrival time probability density function
used to determine \rxte\ times-of-arrival (see \S\ref{sec:ML}).  The
peak for the first Gaussian was explicitly placed at zero. The
listed values for flux indicate the integrated area of
each Gaussian such that the total pulsed signal ``flux'' from the
pulsar equals one.  The DC flux describes the integrated
background level over the full pulse profile.}
\end{deluxetable}
\end{center}

\begin{center}
\begin{deluxetable}{lc}
\tablecaption{Phase-coherent Timing Parameters for \psr\ 
\label{table:parameters}}
\tablewidth{0pt}
\startdata
\hline
\multicolumn{2}{c}{First phase-coherent solution \tablenotemark{a}}
\\ \hline \hline
Dates (Modified Julian Day)        & 52327 -- 52538 \\
Dates                              & 2002 Feb 22 -- 2002 Sep 21 \\
Number of TOAs                     & 78 \\
Epoch (Modified Julian Day)        & 52345.0 \\
$\nu$ (Hz)                         &  15.2238557657(5)\\
$\dot{\nu}$ ($10^{-11}$~s$^{-2}$)  & $-$4.49522(2)  \\
$\ddot{\nu}$ ($10^{-21}$~s$^{-3}$) & 2.00(3) \\
RMS residuals with $\ddot{\nu}$ removed (ms) & 1.56\\
Derivatives needed to `whiten'     & 6 \\
\hline
\multicolumn{2}{c}{Second phase-coherent solution \tablenotemark{a}}
\\ \hline \hline
Dates (Modified Julian Day)        & 52571 --  52776 \\
Dates                              & 2002 Oct 24 -- 2003 May 17 \\
Number of TOAs                     & 33 \\
Epoch (Modified Julian Day)        & 52345 \\
$\nu$ (Hz)                         & 15.22386798(2) \\
$\dot{\nu}$ ($10^{-11}$~s$^{-2}$)  & $-$4.5415(1) \\
$\ddot{\nu}$ ($10^{-21}$~s$^{-3}$) & 12.33(4) \\
RMS residuals with $\ddot{\nu}$ removed (ms) & 1.13 \\
Derivatives needed to `whiten'     & 5 \\
\hline
\multicolumn{2}{c}{Third phase-coherent solution \tablenotemark{a}}
\\ \hline \hline
Dates (Modified Julian Day)        & 53063 -- 54669 \\
Dates                              & 2004 Feb 28 -- 2008 Jul 22 \\
Number of TOAs                     & 379  \\
Epoch (Modified Julian Day)        & 54114.46 \\
$\nu$ (Hz)                         & 15.21701089718(2)\\
$\dot{\nu}$ ($10^{-11}$~s$^{-2}$)  & $-$4.48652358(9)  \\
$\ddot{\nu}$ ($10^{-21}$~s$^{-3}$) & 5.85153(5)\\
RMS residuals with $\ddot{\nu}$ removed (ms) & 965  \\
Derivatives needed to `whiten'     & $>$12   \\
\hline
\enddata
\tablenotetext{a}{Figures in parentheses are uncertainties in the last
digits quoted and are the formal $1 \sigma$ uncertainties reported by TEMPO.}
\end{deluxetable}
\end{center}

\begin{center}
\begin{deluxetable}{lcccc}
\tablecaption{Radio - High energy phase offset for `main' pulse\tablenotemark{a}
\label{offset_table} }
\tablewidth{0pt}
\startdata
\hline
Pulsar     & Period&Radio/X-ray&Radio/Gamma-ray& Refs. \\
           &   (ms)&offset\tablenotemark{b}&offset&  \\
	   \hline\hline
B1937+21     & 1.56 & 0.04(1) & --      & 1 \\ 
J0218+4232   & 2.32 & 0.0(1)  & 0.50(5)  & 2, 3\\
B1821$-$24   & 3.05 & 0.00(2) & 0.00(5) & 4, 5 \\
J0613$-$0200 & 3.06 & --      & 0.42(5) & 3 \\
J1614$-$2230 & 3.15 & --      & 0.20(5) & 3\\
J0751+1807   & 3.48 & --      & 0.42(5) & 3 \\
J1744$-$1134 & 4.08 & --      & $-$0.15(5) & 3 \\
J0030+0451   & 4.87 & 0.00(6)\tablenotemark{c} & 0.15(1) & 6, 3 \\
J2124$-$3358 & 4.93 & --      & $-$0.15(5)    & 3 \\
J0437$-$4715 & 5.76 & 0.003(3)\tablenotemark{c}&0.45(5)& 7, 3 \\
J0737$-$3039A& 22.7 & 0.04(6) & --     & 8 \\
Crab         & 33.1 &$-$0.0102(12)\tablenotemark{d}& $-$0.001(2)& 9, 10\\
B1951+32     & 39.5 & --      & 0.16(6) & 11 \\ 
J2229+6114   & 51.6 & --      & 0.50(3) & 5 \\
J0205+6449   & 65.7 & 0.10(1) & 0.08(2)  & This work, 12\\
J1420$-$6048 & 68.2 & $-$0.35(6) &--      & 13 \\ 
Vela         & 89.3 & 0.12(1) &0.1339(7), 0.130(1)& 14, 10, 15 \\
J1028$-$5819 & 91.4 & --      & 0.200(3)  & 16 \\
B1706$-$44   & 102  & 0.0(2)  &0.211(7)& 17, 10 \\ 
J2021+3651   & 104  &  --     &0.165(10), 0.162(14) & 18, 19  \\
B1509$-$58   & 151  & 0.27(1) &0.38(3),0.30(5) & 4, 20, 5 \\
B1055$-$52   & 197  &$-$0.20(5)\tablenotemark{c}&  $-$0.25(4) & 21, 22 \\
B1929+10     & 227  & 0.06(2) &--      & 23 \\ 
B0950+08     & 253  & 0.25(11)&--      & 24 \\ 
B0656+14     & 385  & $-$0.25(5)\tablenotemark{c}&0.26(8)& 21, 25 \\
J1119$-$6127 & 408  & 0.006(6)\tablenotemark{c}& --     & 26 \\ 
B0628$-$28   & 1244 & 0.20(5) & --     & 27 \\
\hline \enddata
\footnotesize
\tablenotetext{a}{The phase offset is presented from -0.5 to 0.5 in
phase, where a positive value of the phase offset indicates that the
radio pulse leads the high-energy pulse in phase.  Where uncertainties are not
specified in the original publication, an estimate is given.}
\tablenotetext{b}{X-ray pulse represents magnetospheric emission except where otherwise noted.}  
\tablenotetext{c}{Thermal X-ray pulsations.} 
\tablenotetext{d}{The phase lag for the Crab pulsar is calculated from the
phase of the main peak of the radio profile. However, the Crab pulsar also
has a precursor to the main radio pulse seen only at low radio frequencies \citep{mh96}.
If the precursor pulse is the true radio pulse, the phase offset between radio and the 
higher energy pulses is $\sim$0.05 in
phase \citep[see e.g.][]{ry95}.} \\
1. \citet{chk+04}, 
2. \citet{khs04},  
3. \citet{fermi_msp}, 
4. \citet{rjm+98}, 
5. \citet{ppp+09}, 
6. \citet{fermi_0030} 
7. \citet{zps+02}, 
8. \citet{cgm+07}, 
9. \citet{rjl04},  
10. \citet{ppp+08}, 
11. \citet{rbd+95}, 
12. \citet{fermi_0205}, 
13. \citet{rrj01},  
14. \citet{hsgd+02}, 
15. \citet{fermi_vela}, 
16. \citet{fermi_1028}, 
17. \citet{ghd02}, 
18. \citet{hcg+08}, 
19. \citet{fermi_2021}, 
20. \citet{khk+99}, 
21. \citet{dcm+05}, 
22. \citet{tbb+99}, 
23. \citet{bkj+06}, 
24. \citet{zp04},   
25. \citet{hks+97}, 
26. \citet{gkc+05}, 
27. \citet{bjk+05}, 
\end{deluxetable}
\end{center}


\begin{thebibliography}{70}
\expandafter\ifx\csname natexlab\endcsname\relax\def\natexlab#1{#1}\fi

\bibitem[{{Abdo} {et~al.}(2009{\natexlab{a}})}]{fermi_msp}
{Abdo}, A.~A. et.~al. 2009{\natexlab{a}}, Science, 1176113

\bibitem[{{Abdo} {et~al.}(2009{\natexlab{b}})}]{fermi_0205}
---. 2009{\natexlab{b}}, \apjl, 699, L102

\bibitem[{{Abdo} {et~al.}(2009{\natexlab{c}})}]{fermi_1028}
---. 2009{\natexlab{d}}, \apjl, 695, L72

\bibitem[{{Abdo} {et~al.}(2009{\natexlab{d}})}]{fermi_vela}
---. 2009{\natexlab{d}}, \apj, 696, 1084

\bibitem[{{Abdo} {et~al.}(2009{\natexlab{e}})}]{fermi_2021}
---. 2009{\natexlab{e}}, \apj, 700, 1059

\bibitem[{{Abdo} {et~al.}(2009{\natexlab{f}})}]{fermi_0030}
---. 2009{\natexlab{f}}, \apj, 699, 1171

\bibitem[{Arzoumanian {et~al.}(1994)Arzoumanian, Nice, Taylor, \&
  Thorsett}]{antt94}
  Arzoumanian, Z., Nice, D.~J., Taylor, J.~H., \& Thorsett, S.~E. 1994, ApJ,
  422, 671
  
\bibitem[{Backer {et~al.}(1997)Backer, Dexter, Zepka, D., Wertheimer, Ray,
  \& Foster}]{bdz+97}
  Backer, D.~C., Dexter, M.~R., Zepka, A., D., N., Wertheimer, D.~J., Ray,
  P.~S., \& Foster, R.~S. 1997, PASP, 109, 61
  
\bibitem[{{Becker} {et~al.}(2005){Becker}, {Jessner}, {Kramer}, {Testa},
  \&   {Howaldt}}]{bjk+05}
  {Becker}, W., {Jessner}, A., {Kramer}, M., {Testa}, V., \& {Howaldt}, C.
  2005, ApJ, 633, 367
  
\bibitem[{{Becker} {et~al.}(2006){Becker}, {Kramer}, {Jessner}, {Taam},
  {Jia}, {Cheng}, {Mignani}, {Pellizzoni}, {de Luca}, {S{\l}owikowska}, \&
  {Caraveo}}]{bkj+06}
  {Becker}, W., et~al. 2006, \apj, 645, 1421
	
\bibitem[{{Bietenholz}(2006){Bietenholz}}]{bie06}
   {Bietenholz}, M.~F. 2006, \apj, 645, 1180

\bibitem[{{Bietenholz} {et~al.}(2001){Bietenholz}, {Kassim}, \&
  {Weiler}}]{bkw01b}
  {Bietenholz}, M.~F., {Kassim}, N.~E., \& {Weiler}, K.~W. 2001, \apj,
  560, 772

\bibitem[{Buccheri {et~al.}(1983){Buccheri}, {Bennett}, {Bignami},
{Bloemen}, {Boriakoff}, {Caraveo}, {Hermsen}, {Kanbach}, {Manchester},
{Masnou}, {Mayer-Hasselwander}, {Ozel}, {Paul}, {Sacco}, {Scarsi},
{Strong}}]{bbb+83}
 {Buccheri}, R., et~al. 1983, ApJ, 128, 245

\bibitem[{Camilo {et~al.}(2002){Camilo}, {Stairs}, {Lorimer}, {Backer},
  {Ransom}, {Klein}, {Wielebinski}, {Kramer}, {McLaughlin}, {Arzoumanian}, 
  \& {M{\"u}ller}}]{csl+02}
   {Camilo}, F., et~al. 2002, ApJ, 571, L41
  
\bibitem[{{Chatterjee} {et~al.}(2007){Chatterjee}, {Gaensler}, {Melatos},
  {Brisken}, \& {Stappers}}]{cgm+07}
  {Chatterjee}, S., {Gaensler}, B.~M., {Melatos}, A., {Brisken}, W.~F., \&
  {Stappers}, B.~W. 2007, \apj, 670, 1301
      
\bibitem[{{Cheng}(1987)}]{cheng87}
  {Cheng}, K.~S. 1987, \apj, 321, 799
     
\bibitem[{Cheng {et~al.}(1986)Cheng, Ho, \& Ruderman}]{chr86b}
  Cheng, K.~S., Ho, C., \& Ruderman, M. 1986, ApJ, 300, 522
      
\bibitem[{Cordes \& Downs(1985)}]{cd85}
  Cordes, J.~M. \& Downs, G.~S. 1985, ApJS, 59, 343
      
\bibitem[{Cordes \& Greenstein(1981)}]{cg81}
  Cordes, J.~M. \& Greenstein, G. 1981, ApJ, 245, 1060
      
\bibitem[{{Cusumano} {et~al.}(2004){Cusumano}, {Hermsen}, {Kramer},
  {Kuiper}, {L{\"o}hmer}, {Massaro}, {Mineo}, {Nicastro}, \& {Stappers}}]{chk+04}
  {Cusumano}, G., et~al. 2004, Nuclear Physics B Proceedings Supplements, 132, 596

\bibitem[{Daugherty \& Harding(1982)}]{dh82}
  Daugherty, J.~K. \& Harding, A.~K. 1982, ApJ, 252, 337


\bibitem[{{De Jager} {et~al.}(1989){De Jager}, {Raubenheimer}, \&
{Swanepoel}}]{jrs89}
  De Jager, O.~C., et~al. 1989, \aap, 221, 180

\bibitem[{{De Luca} {et~al.}(2005){De Luca}, {Caraveo}, {Mereghetti},
  {Negroni}, \& {Bignami}}]{dcm+05}
  {De Luca}, A., {Caraveo}, P.~A., {Mereghetti}, S., {Negroni}, M., \&
  {Bignami}, G.~F. 2005, \apj, 623, 1051
  
\bibitem[{{Dib} {et~al.}(2007){Dib}, {Kaspi}, \& {Gavriil}}]{dkg07}
  {Dib}, R., {Kaspi}, V.~M., \& {Gavriil}, F.~P. 2007, ApJ, 666, 1152
  
\bibitem[{Fesen(1983)}]{fes83}
  Fesen, R.~A. 1983, \apjl, 270, L53
  
\bibitem[{{Gonzalez} {et~al.}(2005){Gonzalez}, {Kaspi}, {Camilo},
  {Gaensler}, \& {Pivovaroff}}]{gkc+05}
  {Gonzalez}, M.~E., {Kaspi}, V.~M., {Camilo}, F., {Gaensler}, B.~M., \&
  {Pivovaroff}, M.~J. 2005, ApJ, 630, 489
    
\bibitem[{{Gotthelf} {et~al.}(2002){Gotthelf}, {Halpern}, \&
  {Dodson}}]{ghd02}
  {Gotthelf}, E.~V., {Halpern}, J.~P., \& {Dodson}, R. 2002, ApJ, 567, L125

\bibitem[{{Halpern} {et~al.}(2008){Halpern}, {Camilo}, {Giuliani}, {Gotthelf},
  {McLaughlin}, {Mukherjee}, {Pellizzoni}, {Ransom}, {Roberts}, \&
  {Tavani}}]{hcg+08}
  {Halpern}, J.~P., et al. 2008, \apjl, 688, L33
	
\bibitem[{{Harding} {et~al.}(2002){Harding}, {Strickman}, {Gwinn}, {Dodson},
  {Moffet}, \& {McCulloch}}]{hsgd+02}
  {Harding}, A.~K., {Strickman}, M.~S., {Gwinn}, C., {Dodson}, R., {Moffet},
  D., \& {McCulloch}, P. 2002, \apj, 576, 376

\bibitem[{{Hermsen} {et~al.}(1997){Hermsen}, {Kuiper}, {Sch{\"o}nfelder},
  {Strong}, {Bennett}, {Much}, {McConnell}, {Ryan}, \& {Carrami{\~n}ana}}]{hks+97}
  {Hermsen}, W., et al. 1997, in ESA Special Publication, Vol. 382, The Transparent
  Universe, ed. C.~{Winkler}, T.~J.-L. {Courvoisier}, \& P.~{Durouchoux}, 287

\bibitem[{Hobbs {et~al.}(2002)Hobbs, Lyne, Joshi, Kramer,
        Stairs, Camilo, Manchester, D'Amico, Possenti, \& Kaspi}]{hlj+02}
    Hobbs, G., et al. 2002, MNRAS, 333, L7

\bibitem[{Hobbs {et~al.}(2004)Hobbs, Lyne, Kramer, Martin, \&
  Jordan}]{hlk+04}
  Hobbs, G., Lyne, A.~G., Kramer, M., Martin, C.~E., \& Jordan, C. 2004, MNRAS,
  353, 1311
  
\bibitem[{{Jahoda} {et~al.}(2006){Jahoda}, {Markwardt}, {Radeva}, {Rots},
  {Stark}, {Swank}, {Strohmayer}, \& {Zhang}}]{jmr+06}
  {Jahoda}, K., {Markwardt}, C.~B., {Radeva}, Y., {Rots}, A.~H., {Stark},
  M.~J., {Swank}, J.~H., {Strohmayer}, T.~E., \& {Zhang}, W. 2006, \apjs, 163, 401
  
\bibitem[{{Janssen} \& {Stappers}(2006)}]{js06}
              {Janssen}, G.~H. \& {Stappers}, B.~W. 2006,
               \aap, 457, 611
						
\bibitem[{Kaspi {et~al.}(2000)Kaspi, Lackey, \& Chakrabarty}]{klc00}
  Kaspi, V.~M., Lackey, J.~R., \& Chakrabarty, D. 2000, ApJ, 537, L31
  
\bibitem[{Kaspi {et~al.}(2001)Kaspi, Roberts, Vasisht, Gotthelf, Pivovaroff,
  \&  Kawai}]{krv+01}
  Kaspi, V.~M., Roberts, M. S.~E., Vasisht, G., Gotthelf, E.~V., Pivovaroff,
  M., \& Kawai, N. 2001, ApJ, 560, 371
  
\bibitem[{{Kramer} {et~al.}(2003){Kramer}, {Lyne}, {Hobbs}, {L{\" o}hmer},
  {Carr}, {Jordan}, \& {Wolszczan}}]{klh+03}
  {Kramer}, M., {Lyne}, A.~G., {Hobbs}, G., {L{\" o}hmer}, O., {Carr}, P.,
  {Jordan}, C., \& {Wolszczan}, A. 2003, ApJ, 593, L31

\bibitem[{{Kuiper} {et~al.}(1999){Kuiper}, {Hermsen}, {Krijger}, {Bennett},
  {Carrami{\~n}ana}, {Sch{\"o}nfelder}, {Bailes}, \& {Manchester}}]{khk+99}
  {Kuiper}, L., {Hermsen}, W., {Krijger}, J.~M., {Bennett}, K.,
  {Carrami{\~n}ana}, A., {Sch{\"o}nfelder}, V., {Bailes}, M., \& {Manchester},
  R.~N. 1999, A\&A, 351, 119
  
\bibitem[{{Kuiper} {et~al.}(2004){Kuiper}, {Hermsen}, \& {Stappers}}]{khs04}
  {Kuiper}, L., {Hermsen}, W., \& {Stappers}, B. 2004, 33, 507
  

\bibitem[{{Livingstone} {et~al.}(2005){Livingstone}, {Kaspi}, {Gavriil}, \&
  {Manchester}}]{lkgm05}
  {Livingstone}, M.~A., {Kaspi}, V.~M., {Gavriil}, F.~P., \& {Manchester},
  R.~N. 2005, \apj, 619, 1046
  
\bibitem[{Lyne {et~al.}(1993)Lyne, Pritchard, \& Graham-Smith}]{lps93}
  Lyne, A.~G., Pritchard, R.~S., \& Graham-Smith, F. 1993, MNRAS, 265, 1003

\bibitem[{Lyne(1996)}]{lyn96}
Lyne, A.~G. 1996, in Pulsars: Problems and Progress, {IAU} Colloquium 160, ed.
  S.~Johnston, M.~A. Walker, \& M.~Bailes (San Francisco: Astronomical Society
    of the Pacific), 73

\bibitem[{Lyne {et~al.}(1996)Lyne, Pritchard, Graham-Smith, \&
  Camilo}]{lpgc96}
  Lyne, A.~G., Pritchard, R.~S., Graham-Smith, F., \& Camilo, F. 1996, Nature,
  381, 497
  
\bibitem[{{Lyne} {et~al.}(2000){Lyne}, {Shemar}, \& {Graham-Smith}}]{lsg00}
{Lyne}, A.~G., {Shemar}, S.~L., \& {Graham-Smith}, F. 2000, MNRAS, 315, 534
  
\bibitem[{Lyne \& Graham-Smith(2005)}]{ls05a}
Lyne, A.~G. \& Graham-Smith, F. 2005, Pulsar Astronomy, 3rd ed. (Cambridge:
  Cambridge University Press)
  
\bibitem[{McKenna \& Lyne(1990)}]{ml90}
  McKenna, J. \& Lyne, A.~G. 1990, Nature, 343, 349
  
\bibitem[{{Moffett \& Hankins}(1996)}]{mh96}
  {Moffett, D. A. and Hankins, T. H. } 1996, ApJ, 468, 779
  
\bibitem[{Murray {et~al.}(2002)Murray, Slane, Seward, Ransom, \&
   Gaensler}]{mss+02}
   Murray, S.~S., Slane, P.~O., Seward, F.~D., Ransom, S.~M., \& Gaensler,
   B.~M. 2002, ApJ, 568, 226

\bibitem[{{Ng} {et~al.}(2007){Ng}, {Romani}, {Brisken}, {Chatterjee}, \&
  {Kramer}}]{nrb+07}
  {Ng}, C.-Y., {Romani}, R.~W., {Brisken}, W.~F., {Chatterjee}, S., \&
  {Kramer}, M. 2007, \apj, 654, 487
  
\bibitem[{{Pellizzoni} {et~al.}(2009{\natexlab{a}})}]{ppp+08}
    {Pellizzoni}, A., et al. 2009{\natexlab{a}}, ApJ, 691, 1618

\bibitem[{{Pellizzoni} {et~al.}(2009{\natexlab{b}}){Pellizzoni} et al.}]{ppp+09}
    {Pellizzoni}, A., et al. 2009{\natexlab{b}}, \apjl, 695, L115

\bibitem[{Ramanamurthy {et~al.}(1995)Ramanamurthy, Bertsch, Dingus,
   Espositio, Fierro, Fichtel, Etienne, Fichtel, Friedlander, Hunter, Kanbach, Kniffen,
   Lin, Lyne, Mattox, Mayer-Hasselwander, Merck, Michelson, von Montigny,
   Mukherjee, Nolan, \& Thompson}]{rbd+95}
   Ramanamurthy, P.~V., et al. 1995, ApJ, 447, L109
    
\bibitem[{{Ransom}(2001)}]{ran01}
  {Ransom}, S.~M. 2001, Ph.D.~Thesis, Harvard University

\bibitem[{{Ransom} {et~al.}(2004){Ransom}, {Camilo}, {Kaspi}, {Slane},
   {Gaensler}, {Gotthelf}, \& {Murray}}]{rck+04}
   {Ransom}, S., {Camilo}, F., {Kaspi}, V., {Slane}, P., {Gaensler}, B.,
   {Gotthelf}, E., \& {Murray}, S. 2004, in AIP Conf. Proc. 714: X-ray
   Timing 2003: Rossi and Beyond, ed. P.~{Kaaret}, F.~K. {Lamb}, \& J.~H. {Swank},
   350
    
\bibitem[{{Roberts} {et~al.}(2001){Roberts}, {Romani}, \&
  {Johnston}}]{rrj01}
  {Roberts}, M.~S.~E., {Romani}, R.~W., \& {Johnston}, S. 2001, ApJ, 561, L187

\bibitem[{Romani \& Yadigaroglu(1995)}]{ry95}
  Romani, R.~W. \& Yadigaroglu, I.-A. 1995, ApJ, 438, 314

\bibitem[{Romani(1996)}]{rom96a}
  Romani, R.~W. 1996, ApJ, 470, 469

\bibitem[{{Rots} {et~al.}(2004){Rots}, {Jahoda}, \& {Lyne}}]{rjl04}
  {Rots}, A.~H., {Jahoda}, K., \& {Lyne}, A.~G. 2004, \apjl, 605, L129

\bibitem[{Rots {et~al.}(1998)Rots, Jahoda, Macomb, Kawai, Saito, Kaspi, Lyne,
  Manchester, Backer, Somer, Marsden, \& Rothschild}]{rjm+98}
  Rots, A.~H., et~al. 1998, ApJ, 501, 749
    
\bibitem[{{Slane} {et~al.}(2004){Slane}, {Helfand}, {van der Swaluw}, \&
   {Murray}}]{shsm04}
   {Slane}, P., {Helfand}, D.~J., {van der Swaluw}, E., \& {Murray},
   S.~S. 2004, ApJ, 616, 403
  
\bibitem[{Slane {et~al.}(2002)Slane, Helfand, \& Murray}]{shm02}
  Slane, P.~O., Helfand, D.~J., \& Murray, S.~S. 2002, ApJ, 571, L45
  
\bibitem[{{Stephenson} \& {Green}(2002)}]{sg02a}
  {Stephenson}, F.~R. \& {Green}, D.~A. 2002, {Historical supernovae and their
  remnants. International series in astronomy and astrophysics, 
  vol. 5. (Oxford: Clarendon Press)}
      
\bibitem[{{Tam} \& {Roberts}(2003)}]{tr03}
  {Tam}, C. \& {Roberts}, M.~S.~E. 2003, \apj, 598, L27

\bibitem[{Taylor(1992)}]{tay92}
Taylor, J.~H. 1992, Philos. Trans. Roy. Soc. London A, 341, 117

\bibitem[{Thompson {et~al.}(1999)Thompson, Bailes, Bertsch, Cordes, D'Amico,
  Esposito, Finley, Hartman, Hermsen, Kanbach, Kaspi, Kniffen,
  Kuiper, Lin, Manchester, Matz, Mayer-Hasselwander, Michelson, Nolan, Ogelman, Pohl,
  Ramanamurthy, Sreekumar, Reimer, Taylor, \& Ulmer}]{tbb+99}
  Thompson, D.~J., et al. 1999, ApJ, 516, 297
    
\bibitem[{Torii {et~al.}(1999)Torii, Tsunemi, Dotani, Mitsuda, Kawai,
  Kinugasa, Saito, \& Shibata}]{ttd+99}
  Torii, K., Tsunemi, H., Dotani, T., Mitsuda, K., Kawai, N., Kinugasa, K.,
  Saito, Y., \& Shibata, S. 1999, ApJ, 523, L69
    
\bibitem[{{Urama} {et~al.}(2006){Urama}, {Link}, \& {Weisberg}}]{ulw06}
  {Urama}, J.~O., {Link}, B., \& {Weisberg}, J.~M. 2006, \mnras, 370, L76
    
\bibitem[{{Yakovlev} {et~al.}(2002){Yakovlev}, {Kaminker}, {Haensel}, \&
   {Gnedin}}]{ykhg02}
   {Yakovlev}, D.~G., {Kaminker}, A.~D., {Haensel}, P., \& {Gnedin},
   O.~Y. 2002, A\&A, 389, L24
  
\bibitem[{{Zavlin} \& {Pavlov}(2004)}]{zp04}
  {Zavlin}, V.~E. \& {Pavlov}, G.~G. 2004, \apj, 616, 452
  
\bibitem[{{Zavlin} {et~al.}(2002){Zavlin}, {Pavlov}, {Sanwal}, {Manchester},
   {Tr{\" u}mper}, {Halpern}, \& {Becker}}]{zps+02}
   {Zavlin}, V.~E., {Pavlov}, G.~G., {Sanwal}, D., {Manchester}, R.~N., 
   {Tr{\"u}mper}, J., {Halpern}, J.~P., \& {Becker}, W. 2002, ApJ, 569, 894

\end{thebibliography}
\end{document}